\documentclass[preprintnumbers,amsmath,amssymb,showpacs]{revtex4}
\usepackage{epsfig}
\usepackage{color}
\usepackage[colorlinks,urlcolor=blue,citecolor=red]{hyperref}
\begin{document}
\setlength{\voffset}{1.0cm}
\title{Solving the U(2)$_L \times$U(2)$_R$ symmetric Nambu--Jona-Lasinio model in 1+1 dimensions}
\author{Michael Thies\footnote{michael.thies@gravity.fau.de}}
\affiliation{Institut f\"ur  Theoretische Physik, Universit\"at Erlangen-N\"urnberg, D-91058, Erlangen, Germany}
\date{\today}
\begin{abstract}
A less well known variant of the Nambu--Jona-Lasinio model with $N_c$ colors and U(2)$_L \times$U(2)$_R$ chiral symmetry is studied in 1+1 dimensions.
Using semi-classical methods appropriate for the large $N_c$ limit, we determine the vacuum manifold, the meson spectrum, massless and massive
multi-fermion bound states and the phase diagram as a function of temperature, chemical potential and isospin chemical potential. An important tool
to understand soliton dynamics is the generalization of the time-dependent Hartree-Fock approach to two flavors along the lines recently
developed by Takahashi in the context of unconventional fermionic superfluids and superconductors. 
\end{abstract}
\maketitle
%<<<<<<<<<<<<<<<<<<<<<<<<<<<<<<<<<<<<<<<<<<<<<<<<<<<<<<<<<<<<<<<<<<<<<<<<<<<<<<<<<<<<<<<<<<<< <<<<<<<<<<<<<<<<<<<<<<<<<<<<<
%<<<<<<<<<<<<<<<<<<<<<<<<<<<<<<<<<<<<<<<<<<<<<<<<<<<<<<<<<<<<<<<<<<<<<<<<<<<<<<<<<<<<<<<<<<<<<<<<<<<<<<<<<<<<<<<<<<<<<<<<<<
\section{Introduction}
\label{sect1}
%<<<<<<<<<<<<<<<<<<<<<<<<<<<<<<<<<<<<<<<<<<<<<<<<<<<<<<<<<<<<<<<<<<<<<<<<<<<<<<<<<<<<<<<<<<<<<<<<<<<<<<<<<<<<<<<<<<<<<<<<<<
%<<<<<<<<<<<<<<<<<<<<<<<<<<<<<<<<<<<<<<<<<<<<<<<<<<<<<<<<<<<<<<<<<<<<<<<<<<<<<<<<<<<<<<<<<<<<<<<<<<<<<<<<<<<<<<<<<<<<<<<<<<
The two best studied variants of the massless Gross-Neveu (GN) model \cite{L1} differ by their chiral symmetry groups. The first model has the Lagrangian
\begin{equation}
%{\cal L} = \bar{\psi} i \partial \!\!\!/ \psi + \frac{g^2}{2}  (\bar{\psi}\psi)^2 \qquad({\rm Z}_2- {\rm GN})
{\cal L} = \bar{\psi} i \partial \!\!\!/ \psi + \frac{g^2}{2}  (\bar{\psi}\psi)^2 \qquad({\rm  Z}_2{\text -}{\rm GN}).
\label{A1}
\end{equation}  
Throughout this paper, we are always in 1+1 dimensions and suppress contracted ``color" labels on the fermion bilinears (1...$N_c$).
The Lagrangian (\ref{A1}) is invariant under independent sign flips of left- and right-handed Dirac fields, corresponding to the discrete 
chiral group  Z$_{2,L} \times $Z$_{2,R}$. The second model has a continuous U(1)$_L \times $U(1)$_R$
chiral symmetry and is often referred to as two-dimensional Nambu--Jona-Lasinio (NJL$_2$) model \cite{L2},
\begin{equation}
{\cal L} = \bar{\psi} i \partial \!\!\!/ \psi + \frac{g^2}{2} \left[ (\bar{\psi}\psi)^2 + (\bar{\psi} i \gamma_5 \psi)^2  \right] \qquad ({\rm U(1)}{\text -}{\rm NJL}).
\label{A2}
\end{equation} 
Here, phases of left- and right-handed fermion fields can be rotated independently. As a phenomenological model, the NJL model in 3+1 dimensions
is in general considered with $N_c=3$, but two additional flavors ($N_f=2$) corresponding to isospin in strong interaction physics \cite{L3}. This leads to
the Lagrangian with SU(2)$_L \times $SU(2)$_R$ chiral symmetry
\begin{equation}
{\cal L} = \bar{\psi} i \partial \!\!\!/ \psi + \frac{g^2}{2} \left[ (\bar{\psi}\psi)^2 + (\bar{\psi}i \gamma_5 \vec{\tau} \psi)^2 \right]  \qquad ({\rm SU(2)}{\text -}{\rm NJL}).
\label{A3}
\end{equation}
The same theory in 1+1 dimensions is also exactly solvable in the large $N_c$ limit, see the recent paper \cite{L4}. 
As shown there and noticed before in the context of the phase diagram \cite{L5}, the physics is closer to that of the GN model (\ref{A1}) than
to that of the NJL$_2$ model (\ref{A2}). This reflects the following group-theoretical fact: The SU(2) chiral symmetry also entails the discrete 
symmetry of the one-flavor GN model through the center of the group SU(2), but obviously not the full U(1) symmetry. If Lagrangian (\ref{A3}) is the generalization of Lagrangian 
(\ref{A1}) to two flavors, it is not too hard to identify the generalization of Lagrangian (\ref{A2}) to two flavors: It is the NJL-type
model with U(2)$_L \times$U(2)$_R$ chiral symmetry and Lagrangian
\begin{equation}
{\cal L} = \bar{\psi} i \partial \!\!\!/ \psi + \frac{g^2}{2} \left[ (\bar{\psi}\psi)^2 + (\bar{\psi} \vec{\tau} \psi)^2
+ (\bar{\psi} i \gamma_5 \psi)^2 + (\bar{\psi}i \gamma_5 \vec{\tau} \psi)^2 \right]       \qquad ({\rm U(2)}{\text -}{\rm NJL}).
\label{A4}
\end{equation}
In which sense model (\ref{A4}) is the generalization of model (\ref{A2}) and model (\ref{A3}) the generalization of model (\ref{A1}) will be made more
precise below, after having the solutions of all four models at our disposal.

The U(2)-NJL model has been considered in 3+1 dimensions before \cite{L6,L7,L8}. In 1+1 dimensions, it has been explored using non-Abelian bosonization
at finite $N_c$ \cite{L8a}. 
When going through the four models (\ref{A1})--(\ref{A4}), the complexity increases significantly, as evidenced by the number of (real) bilinear condensates
which is doubled at each step ($1\to 2 \to 4 \to 8$). Nevertheless, according to the
experience with the one-flavor models, the solution of the U(2)-NJL model is expected to be simpler than that of the SU(2)-NJL model. 

Model (\ref{A4})  is also interesting for
another reason: Recently, Takahashi has generalized solutions of the Bogoliubov-de Gennes (BdG) equation to multicomponent form in the context of unconventional Fermi
superfluids \cite{L9,L10}. This work has been
found useful for solving twisted kink dynamics in the SU(2) symmetric model \cite{L4}. It was noticed there that the elementary kink of Takahashi, the 
simplest multi-fermion bound state, does not exist in the SU(2) model, since it is not charge conjugation invariant. For the same  
reason the kink of Shei in the NJL$_2$ model \cite{L11} does not appear in isolation in the GN model, only as a constituent of kink-antikink bound states
(Dashen-Hasslacher-Neveu baryon \cite{L12}). It is then plausible, and will indeed be confirmed below, that Takahashi's kink
will appear in the model (\ref{A4}) as a physical state.

The present paper deals mostly with Lagrangian (\ref{A4}), but cross references to the other variants of the GN model will be frequent. It is
therefore necessary to adopt a simple naming of the four distinct models. Since our emphasis is on the chiral symmetry group, we will
refer in the present work to the models (\ref{A1}) -- (\ref{A4})  as Z$_2$-GN, U(1)-NJL, SU(2)-NJL and U(2)-NJL models, respectively,
as already indicated in the equations. We also omit the subscript 2 for the number of dimensions from now on.

This paper is organized as follows. Sect.~\ref{sect2} develops the mean field theory of the U(2)-NJL model, i.e., the Hartree-Fock (HF) and time-dependent Hartree-Fock (TDHF) 
approaches. In Sect.~\ref{sect3}, spontaneous chiral symmetry 
breaking, the dynamical fermion mass and the vacuum manifold are discussed. The meson spectrum is the topic of Sect.~\ref{sect4}, where the relativistic
random phase approximation (RPA) is invoked to analyze small fluctuations around the HF vacuum. Massless hadrons, chiral spiral type condensates and the phase diagram
as a function of $(T, \mu, \mu_3)$ are all related to the chiral anomaly and the fact that baryon number has a topological interpretation, as discussed in Sect.~\ref{sect5}.
In Sect.~\ref{sect6} we generalize our previous solution of the TDHF equations from the U(1)-NJL model to the two-flavor case, rederiving Takahashi's recent work on the BdG
equation in a slightly different language. This has been done to make sure that the formalism from condensed matter physics really applies to the
relativistic quantum field case at hand, and to facilitate comparison with previous works on other variants of GN or NJL models \cite{L13,L14,L15}. A few simple applications 
to soliton problems with one and two bound states are given in Sects.~\ref{sect7} and \ref{sect8}. We finish with a short concluding section, Sect.~\ref{sect9}.

%<<<<<<<<<<<<<<<<<<<<<<<<<<<<<<<<<<<<<<<<<<<<<<<<<<<<<<<<<<<<<<<<<<<<<<<<<<<<<<<<<<<<<<<<<<<< <<<<<<<<<<<<<<<<<<<<<<<<<<<<<
%<<<<<<<<<<<<<<<<<<<<<<<<<<<<<<<<<<<<<<<<<<<<<<<<<<<<<<<<<<<<<<<<<<<<<<<<<<<<<<<<<<<<<<<<<<<<<<<<<<<<<<<<<<<<<<<<<<<<<<<<<<
\section{Mean field theory}
\label{sect2}
%<<<<<<<<<<<<<<<<<<<<<<<<<<<<<<<<<<<<<<<<<<<<<<<<<<<<<<<<<<<<<<<<<<<<<<<<<<<<<<<<<<<<<<<<<<<<<<<<<<<<<<<<<<<<<<<<<<<<<<<<<<
%<<<<<<<<<<<<<<<<<<<<<<<<<<<<<<<<<<<<<<<<<<<<<<<<<<<<<<<<<<<<<<<<<<<<<<<<<<<<<<<<<<<<<<<<<<<<<<<<<<<<<<<<<<<<<<<<<<<<<<<<<<

The Lagrangian (\ref{A4})  has the U(2)$_L \times$U(2)$_R$ chiral symmetry
\begin{equation}
\psi_L \to e^{i\alpha_0} e^{i \vec{\alpha} \vec{\tau}} \psi_L, \quad  \psi_R \to e^{i\beta_0} e^{i \vec{\beta} \vec{\tau}} \psi_R, \quad \psi_{R,L} = \frac{1\pm \gamma_5}{2} \psi,
\label{B1}
\end{equation}
giving rise to 8 conserved Noether currents
\begin{equation}
j^{\mu} = \bar{\psi} \gamma^{\mu} \psi, \quad j_5^{\mu} = \bar{\psi} \gamma^{\mu}\gamma_5 \psi, \quad
j_a^{\mu} = \bar{\psi} \gamma^{\mu} \tau_a \psi, \quad j_{5,a}^{\mu} = \bar{\psi} \gamma^{\mu}\gamma_5 \tau_a \psi.
\label{B2}
\end{equation}
In 1+1 dimensions, vector and axial vector currents are not independent, but satisfy
\begin{eqnarray}
j_5^0 & = & j^1, \quad j_5^1 = j^0,
\nonumber \\
j_{5,a}^0 & = & j_a^1, \quad j_{5,a}^1 = j_a^0.
\label{B3}
\end{eqnarray}
Adding and subtracting the conservation laws $\partial_{\mu}j^{\mu}=0, \partial_{\mu} j_5^{\mu}=0$, and introducing light cone coordinates
\begin{equation}
z=x-t, \quad \bar{z} = x+t, \quad \partial_0 = \bar{\partial}-\partial, \quad \partial_1 = \bar{\partial}+\partial,
\label{B4}
\end{equation}
one finds
\begin{equation}
\bar{\partial} \psi_R^{\dagger} \psi_R=0, \quad \partial \psi_L^{\dagger} \psi_L = 0.
\label{B5}
\end{equation}
If we take the expectation value of these equations in an arbitrary state, we conclude that the right-handed density $\rho_R = \langle \psi_R^{\dagger}\psi_R \rangle$
depends only on $z$, the left-handed density $\rho_L = \langle \psi_L^{\dagger}\psi_L \rangle$ only on $\bar{z}$, i.e., they can only move with the velocity of light
to the right or to the left (or be constant). In a localized, massive state like a solitonic bound state or breather, these densities must therefore vanish identically. The same
argument goes through for the isovector currents and densities, since
\begin{equation}
\bar{\partial} \psi_R^{\dagger} \tau_a \psi_R=0, \quad \partial \psi_L^{\dagger}\tau_a  \psi_L = 0.
\label{B6}
\end{equation}
Hence we anticipate that all densities and current densities must be zero inside an arbitrary
soliton or multi-soliton state, a strong constraint on multi-fermion states following from chiral symmetry.

The basic equation in the large $N_c$ limit is the relativistic version of the time dependent Hartree-Fock (TDHF) equation. In the present case of Lagrangian (\ref{A4}), it reads
\begin{equation}
\left[ i \gamma^{\mu} \partial_{\mu} - (S_0 + \vec{S}\vec{\tau}\,) - i \gamma_5 (P_0 + \vec{P}\vec{\tau}\,) \right] \psi = 0
\label{B7}
\end{equation}
together with the self-consistency conditions
\begin{eqnarray}
S_0 & = & - g^2 \langle \bar{\psi} \psi \rangle, \quad \vec{S} = - g^2 \langle \bar{\psi} \vec{\tau} \psi \rangle,
\nonumber \\
P_0 & = & - g^2 \langle \bar{\psi} i \gamma_5 \psi \rangle, \quad \vec{P} = - g^2 \langle \bar{\psi}i \gamma_5 \vec{\tau} \psi \rangle.
\label{B8}
\end{eqnarray}
We choose a chiral representation of the Dirac matrices (diagonal $\gamma_5$)
\begin{equation}
\gamma^0 = \sigma_1, \quad \gamma^1 = i \sigma_2, \quad \gamma_5 = \gamma^0 \gamma^1 = - \sigma_3.
\label{B9}
\end{equation}
The TDHF equation in Hamiltonian form then assumes the form
\begin{equation}
i \partial_t \left( \begin{array}{c} \psi_1 \\ \psi_2 \end{array} \right) = 
\left( \begin{array}{cc} i \partial_x & \Delta^{\dagger} \\ \Delta & -i \partial_x \end{array} \right) 
\left( \begin{array}{c} \psi_1 \\ \psi_2 \end{array} \right).
\label{B10}
\end{equation}
Here, the Hamiltonian has been written in 2$\times$2-block-form, $\psi_1=\psi_L$ and $\psi_2=\psi_R$ are 2d isospinors, and
$\Delta$ is the 2$\times$2 matrix
\begin{equation}
\Delta = (S_0-iP_0) + (\vec{S}-i \vec{P})\vec{\tau} := \Delta_0 + \vec{\Delta}\vec{\tau}.
\label{B11}
\end{equation}
The self-consistency conditions (\ref{B8}) are particularly concise in the 2$\times$2 matrix form
\begin{equation}
\Delta = - 2 N g^2 \sum^{\rm occ} \psi_2 \psi_1^{\dagger}= -2N g^2 \sum^{\rm occ} \left( \begin{array}{cc} \psi_{2,1}\psi_{1,1}^* & \psi_{2,1} \psi_{1,2}^* \\
\psi_{2,2}  \psi_{1,1}^*  & \psi_{2,2} \psi_{1,2}^* \end{array} \right) \quad (N=2N_c).
\label{B12}
\end{equation}
In light cone coordinates (\ref{B4})
the covariant form of the TDHF equation takes on the form
\begin{equation}
2 i \bar{\partial} \psi_2 = \Delta \psi_1, \quad 2i \partial \psi_1 = - \Delta^{\dagger} \psi_2, 
\label{B13}
\end{equation}
a 2-component generalization of the equations familiar from the one-flavor case.
Not only the Dirac-TDHF equation (\ref{B13}), but also the self-consistency condition (\ref{B12}) are manifestly preserved under chiral transformations
\begin{equation}
\psi_1 \to U_L \psi_1, \quad \psi_2 \to U_R \psi_2, \quad \Delta \to U_R \Delta U_L^{\dagger}, \qquad  U_{R,L} \in {\rm U(2)}. 
\label{B14}
\end{equation}
That is to say that two solutions which differ only by a chiral transformation have to be considered as being physically indistinguishable. 

%<<<<<<<<<<<<<<<<<<<<<<<<<<<<<<<<<<<<<<<<<<<<<<<<<<<<<<<<<<<<<<<<<<<<<<<<<<<<<<<<<<<<<<<<<<<< <<<<<<<<<<<<<<<<<<<<<<<<<<<<<
%<<<<<<<<<<<<<<<<<<<<<<<<<<<<<<<<<<<<<<<<<<<<<<<<<<<<<<<<<<<<<<<<<<<<<<<<<<<<<<<<<<<<<<<<<<<<<<<<<<<<<<<<<<<<<<<<<<<<<<<<<<
\section{Vacuum and dynamical fermion mass}
\label{sect3}
%<<<<<<<<<<<<<<<<<<<<<<<<<<<<<<<<<<<<<<<<<<<<<<<<<<<<<<<<<<<<<<<<<<<<<<<<<<<<<<<<<<<<<<<<<<<<<<<<<<<<<<<<<<<<<<<<<<<<<<<<<<
%<<<<<<<<<<<<<<<<<<<<<<<<<<<<<<<<<<<<<<<<<<<<<<<<<<<<<<<<<<<<<<<<<<<<<<<<<<<<<<<<<<<<<<<<<<<<<<<<<<<<<<<<<<<<<<<<<<<<<<<<<<

If chiral symmetry is spontaneously broken but Lorentz invariance is preserved, the vacuum is characterized by a space-time independent mean field $\Delta_{\rm vac}$.
In order to decide whether this happens, we diagonalize the HF Hamiltonian (a $4 \times 4$ matrix) with homogeneous $\Delta_{\rm vac}$, Eq.~(\ref{B11}), in momentum space,
\begin{equation}
H = \left( \begin{array}{cc} -k & \Delta_{\rm vac}^{\dagger} \\ \Delta_{\rm vac} & k \end{array} \right) .
\label{C1}
\end{equation}
The 4 eigenvalues are
\begin{equation}
\pm \sqrt{m_1^2 + k^2}, \quad \pm \sqrt{m_2^2+ k^2},
\label{C2}
\end{equation}
with dynamical fermion masses
\begin{eqnarray}
m_{1,2}^2 & = & S_0^2 + \vec{S}^{\,2} + P_0^2 + \vec{P}^{\,2} \pm 2 \sqrt{Z},
\nonumber \\
Z & = &  (S_0\vec{S} + P_0 \vec{P}\,)^2 + (\vec{S} \times \vec{P}\,)^2.  
\label{C3}
\end{eqnarray}
Next we minimize the vacuum energy density, following closely the corresponding steps in the Z$_2$-GN model (see e.g. \cite{L16}),
\begin{equation}
{\cal E}_{\rm vac} = {\cal E}_{\rm sp}(m_1) + {\cal E}_{\rm sp}(m_2)+ {\cal E}_{\rm dc},
\label{C4}
\end{equation}
where ${\cal E}_{\rm sp}(m)$ is the single particle vacuum energy density of the Z$_2$-GN model,
\begin{eqnarray}
{\cal E}_{\rm sp}(m) & = & -N_c \int_{-\Lambda/2}^{\Lambda/2} \frac{dk}{2\pi} \sqrt{k^2+m^2} 
\nonumber \\
& = & - N_c \left[ \frac{\Lambda^2}{8\pi} - \frac{m^2}{4\pi} \left( \ln \frac{m^2}{\Lambda^2} -1 \right)    \right] ,
\label{C5}
\end{eqnarray}
whereas ${\cal E}_{\rm dc}$ denotes the double counting correction of the interaction energy density characteristic for the HF approach,
\begin{equation}
{\cal E}_{\rm dc} = \frac{m_1^2+m_2^2}{4g^2}.
\label{C5a}
\end{equation}
Minimizing ${\cal E}_{\rm vac}$ with respect to $m_1,m_2$ yields the two conditions
\begin{equation}
0 = 1 + \frac{N_c g^2}{\pi} \ln \frac{m_i^2}{\Lambda^2} \quad (i=1,2).
\label{C6}
\end{equation}
Not surprisingly, the minimum is at $m_1=m_2:=m$, and we recover the gap equation of the Z$_2$-GN model with $2N_c$ flavors.
The renormalized vacuum energy density is accordingly
\begin{equation}
{\cal E}_{\rm vac} = - N_c \frac{m^2}{2\pi}.
\label{C7}
\end{equation}
What is the vacuum manifold? The condition $m_1=m_2$ implies  $Z=0$, Eq.~(\ref{C3}), or
\begin{equation}
\vec{S} = - \frac{P_0}{S_0} \vec{P}.
\label{C8}
\end{equation}
The dynamical mass then becomes
\begin{equation}
m^2 = S_0^2 + \vec{S}^2 + P_0^2 + \vec{P}^2 = \frac{(S_0^2+P_0^2)(S_0^2+\vec{P}^{\,2})}{S_0^2}.
\label{C9}
\end{equation}
The vacuum potential matrix $\Delta_{\rm vac}$ [see (\ref{B11})] is given by
\begin{equation}
\Delta_{\rm vac} = \frac{(S_0-iP_0)(S_0-i\vec{P}\vec{\tau})}{S_0}
\label{C10}
\end{equation}
and satisfies
\begin{equation}
\Delta_{\rm vac} \Delta_{\rm vac}^{\dagger}= m^2, \quad {\rm det}\, \Delta_{\rm vac} = m^2 \frac{(S_0-iP_0)^2}{S_0^2+P_0^2}.
\label{C11}
\end{equation}
Choosing units where $m=1$, we first divide $\Delta_{\rm vac}$ by $\sqrt{m^2}$ from Eq.~(\ref{C9}) to get the U(2) matrix
\begin{equation}
\Delta_{\rm vac} = \frac{S_0-iP_0}{\sqrt{S_0^2+P_0^2}} \frac{S_0-i \vec{P}\vec{\tau}}{\sqrt{S_0^2 + \vec{P}^{\,2}}} \in {\rm U(2)}
\label{C12}
\end{equation}
where the first factor belongs to U(1), the second to SU(2). The condition $m=1$ implies furthermore that 
\begin{equation}
S_0^2 = (S_0^2+P_0^2)(S_0^2+\vec{P}^{\,2})
\label{C13}
\end{equation}
Since a U(2) matrix has 4 real parameters, but $\Delta$ in (\ref{C12}) is parametrized by 5 real numbers ($S_0,P_0,\vec{P}$), one condition
is indeed necessary. Alternatively, it is obvious that expression (\ref{C12}) depends only on the 4 parameters $P_0/S_0, \vec{P}/S_0$. Thus the vacuum manifold
is U(2) or $S^1 \times S^3$. In cases where only a single vacuum is involved, the simplest choice is $\Delta_{\rm vac}=1$, which can always be achieved by
a chiral transformation. For twisted field configurations, we shall use $\Delta_{\rm vac}=1$ for $x\to - \infty$, but we then need the general expression (\ref{C12})
for $x \to \infty$.

%<<<<<<<<<<<<<<<<<<<<<<<<<<<<<<<<<<<<<<<<<<<<<<<<<<<<<<<<<<<<<<<<<<<<<<<<<<<<<<<<<<<<<<<<<<<< <<<<<<<<<<<<<<<<<<<<<<<<<<<<<
%<<<<<<<<<<<<<<<<<<<<<<<<<<<<<<<<<<<<<<<<<<<<<<<<<<<<<<<<<<<<<<<<<<<<<<<<<<<<<<<<<<<<<<<<<<<<<<<<<<<<<<<<<<<<<<<<<<<<<<<<<<
\section{Meson spectrum}
\label{sect4}
%<<<<<<<<<<<<<<<<<<<<<<<<<<<<<<<<<<<<<<<<<<<<<<<<<<<<<<<<<<<<<<<<<<<<<<<<<<<<<<<<<<<<<<<<<<<<<<<<<<<<<<<<<<<<<<<<<<<<<<<<<<
%<<<<<<<<<<<<<<<<<<<<<<<<<<<<<<<<<<<<<<<<<<<<<<<<<<<<<<<<<<<<<<<<<<<<<<<<<<<<<<<<<<<<<<<<<<<<<<<<<<<<<<<<<<<<<<<<<<<<<<<<<<

In order to derive the spectrum of fermion-antifermion bound states (mesons), one has to go beyond the HF approximation and consider small fluctuations
around the HF vacuum. The appropriate tool is the relativistic RPA. Although the technicalities are fairly involved,
we can be brief here since one can follow almost literally the corresponding caculation in the Z$_2$-GN model, described in more detail
in Refs.~\cite{L16,L17}. One starts from the equation of motion for the color singlet bilinear operator
\begin{equation}
Q_{\alpha \beta}(x,y) = \frac{1}{N_c} \sum_{i=1}^{N_c} \psi_{i,\beta}^{\dagger}(y) \psi_{i, \alpha}(x).
\label{D1}
\end{equation}  
For earlier applications of this method, mainly to two-dimensional quantum chromodynamics, see also Refs.~\cite{L17a,L17b,L17c}.
In the present case, $\alpha, \beta$ are combined Dirac- and isospin indices ranging from 1 to 4. The $4 \times 4$ matrix operator $Q(x,y)$ satisfies
the equation of motion
\begin{eqnarray}
i \partial_t Q(x,y) & = & -i \left\{ \partial_y Q(x,y) \gamma_5 + \gamma_5 \partial_x Q(x,y) \right\}
\nonumber \\
& - & N_c g^2 \sum_{n=1}^4 \left\{ {\rm Tr} \left[ O_n Q(x,x) \right] O_n Q(x,y) - Q(x,y) O_n {\rm Tr} \left[ O_n Q(y,y)\right] \right\}
\label{D2}
\end{eqnarray}
where
\begin{equation}
O_1 = \gamma^0, \quad O_2 = i \gamma^1, \quad O_3 = \gamma^0 \vec{\tau}, \quad O_4 = i \gamma^1 \vec{\tau}.
\label{D3}
\end{equation}
We have dropped terms irrelevant in the large $N_c$ limit. In the case of isovector operators $O_3, O_4$, scalar products are implied in
Eq.~(\ref{D2}). In the Z$_2$-GN model, only the $O_1$-term was present. The next steps will not be shown in detail, but only enumerated:
\begin{enumerate}
\item Expand the bilinear operator around the vacuum expectation value
\begin{equation}
Q(x,y) = \rho(x-y) + \frac{1}{\sqrt{N_c}}  \tilde{Q}(x,y). 
\label{D4}
\end{equation}
\item Linearize the equation of motion in the fluctuation part $\tilde{Q}(x,y)$.
\item Choose the vacuum $\Delta_{\rm vac}=1$ without loss of generality.
\item Transform $\tilde{Q}$ and $\rho$ to momentum space.
\item Expand $\tilde{Q}$ into vacuum spinors $u,v$. In the large $N_c$ limit, only particle-hole components are important,
\begin{equation}
\tilde{Q}(k',k) = u(k') v^{\dagger}(k) \tilde{Q}_{12}(k',k) + v(k') u^{\dagger}(k) \tilde{Q}_{21}(k',k).
\label{D5}
\end{equation}
This expansion is related to the Dirac indices only, so that $\tilde{Q}_{12}, \tilde{Q}_{21}$ are still 2$\times$2 matrices consisting of isoscalar and 
isovector pieces,
\begin{equation}
\tilde{Q}_{ij} = \tilde{Q}_{ij}^{(0)} + \tau_a \tilde{Q}_{ij}^{(a)} \quad (ij=12,21). 
\label{D6}
\end{equation}
\item Sandwich the equation of motion between the vacuum and one-meson states with momentum $P$,
\begin{eqnarray}
\langle P | \tilde{Q}_{21}(k',k) |{\rm vac}\rangle & = &  2\pi \delta(P-k+k') X(P,k),
\nonumber \\
\langle P | \tilde{Q}_{12}(k',k) |{\rm vac}\rangle & = &  2\pi \delta(P-k+k') Y(P,k).
\label{D7}
\end{eqnarray}
\end{enumerate}
The RPA-amplitudes $X,Y$ then satisfy coupled linear integral equations of the standard RPA form. 
What one finds is that the scalar and pseudoscalar mesons decouple, as do the 4 different isospin
components (isoscalar and isovector). The kernel of the integral equations is 1-term separable, so
that there can be at most one meson bound state per channel. 
Owing to the separability one can easily obtain an eigenvalue condition from the RPA equations.
All 4 scalar mesons satisfy the same equation leading to the eigenvalue condition
\begin{equation}
1 = N_c g^2 \int \frac{dk}{2\pi} \left( \frac{1}{E(k-P)} + \frac{1}{E(k)} \right) \left\{ \frac{4m^2+P^2 - E^2(k-P,k)}{{\cal E}^2(P) - E^2(k-P,k)} \right\}
\label{D8}
\end{equation}
identical to the one of the Z$_2$-GN model with $2N_c$ flavors. We use the notation
\begin{equation}
E(k)=\sqrt{k^2+m^2}, \quad E(k-P,k) = E(k-P)+E(k),
\label{D9}
\end{equation} 
and ${\cal E}(P)=\sqrt{{\cal M}^2+P^2}$ is the meson energy. For meson mass ${\cal M}=2m$, the factor in curly brackets in Eq.~(\ref{D8}) becomes 1 and the whole equation
reduces to the vacuum gap equation,
hence the 4 scalar mesons all have the same mass ${\cal M}=2m$. The 4 pseudoscalar mesons also satisfy identical equations,
leading to the different eigenvalue condition
\begin{equation}
1 = N_c g^2 \int \frac{dk}{2\pi} \left( \frac{1}{E(k-P)} + \frac{1}{E(k)} \right) \left\{ \frac{P^2 - E^2(k-P,k)}{{\cal E}^2(P) - E^2(k-P,k)} \right\}
\label{D10}
\end{equation}
from which we can read off
a vanishing meson mass ${\cal M}=0$. These are the would-be Goldstone bosons, reflecting the 4 flat directions (i.e., the dimension of the 
vacuum manifold). Note that covariance is manifest in the 
relativistic RPA. Since the meson spectrum has not been discussed in Ref.~\cite{L4}, let us mention that a corresponding calculation for the SU(2)-NJL
model would have yielded one massive scalar/isoscalar (${\cal M}=2m$) and three massless pseudoscalar/isovector mesons.

%<<<<<<<<<<<<<<<<<<<<<<<<<<<<<<<<<<<<<<<<<<<<<<<<<<<<<<<<<<<<<<<<<<<<<<<<<<<<<<<<<<<<<<<<<<<< <<<<<<<<<<<<<<<<<<<<<<<<<<<<<
%<<<<<<<<<<<<<<<<<<<<<<<<<<<<<<<<<<<<<<<<<<<<<<<<<<<<<<<<<<<<<<<<<<<<<<<<<<<<<<<<<<<<<<<<<<<<<<<<<<<<<<<<<<<<<<<<<<<<<<<<<<
\section{Massless hadrons, chiral spiral, phase diagram}
\label{sect5}
%<<<<<<<<<<<<<<<<<<<<<<<<<<<<<<<<<<<<<<<<<<<<<<<<<<<<<<<<<<<<<<<<<<<<<<<<<<<<<<<<<<<<<<<<<<<<<<<<<<<<<<<<<<<<<<<<<<<<<<<<<<
%<<<<<<<<<<<<<<<<<<<<<<<<<<<<<<<<<<<<<<<<<<<<<<<<<<<<<<<<<<<<<<<<<<<<<<<<<<<<<<<<<<<<<<<<<<<<<<<<<<<<<<<<<<<<<<<<<<<<<<<<<<

For the present purpose, it is sufficient to 
consider the case where the internal rotation axis is ``frozen", say in the 3-direction,
\begin{equation}
\Delta = \Delta_0 + \Delta_3 \tau_3.
\label{E1}
\end{equation}
Then the two isospin channels decouple, each one corresponding to a U(1)-NJL model with $N_c$ flavors and mean field
\begin{equation}
S-iP=\Delta_0 \pm \Delta_3=(S_0 \pm S_3)-i(P_0\pm P_3).
\label{E2}
\end{equation}
In the same way as Z$_2$-GN solutions are particular solutions of the SU(2)-NJL model \cite{L4}, we thus find that U(1)-NJL solutions
are particular solutions of the U(2)-NJL model. The TDHF equations are evidently satisfied, as are the self-consistency conditions.
The vacuum can be chosen as $\Delta_{\rm vac}=1$. To describe dense matter, we combine two U(1)-chiral spirals \cite{L18,L19} in the two isospin channels to the following 
transformation of the vacuum spinors
\begin{equation}
\psi \to e^{iax\gamma_5} e^{ibx\tau_3 \gamma_5} \psi.
\label{E3}
\end{equation}
The two resulting chiral spirals have same radius ($m=1$) but different pitches,
\begin{eqnarray}
b_{\rm eff} = a \pm b, \quad S-iP=e^{2ib_{\rm eff} x}, \quad \rho= \frac{N_c b_{\rm eff}}{\pi}, \quad {\cal E} = \frac{N_c b_{\rm eff}^2}{2\pi}
\label{E4}
\end{eqnarray}
($+$ sign for isospin up, $-$ sign for isospin down). In this manner matter with arbitrary $\rho,\rho_3$ can be described most efficiently,
\begin{equation}
\Delta = e^{2i(a+b\tau_3)x}, \quad \rho= \frac{2N_c a}{\pi}, \quad \rho_3= \frac{2 N_c b}{\pi}, \quad {\cal E} = \frac{N_c(a^2+b^2)}{\pi}.
\label{E5}
\end{equation}
In the low density limit we find massless, delocalized  baryons with different isospin content.

\begin{figure}[h]
\begin{center}
\epsfig{file=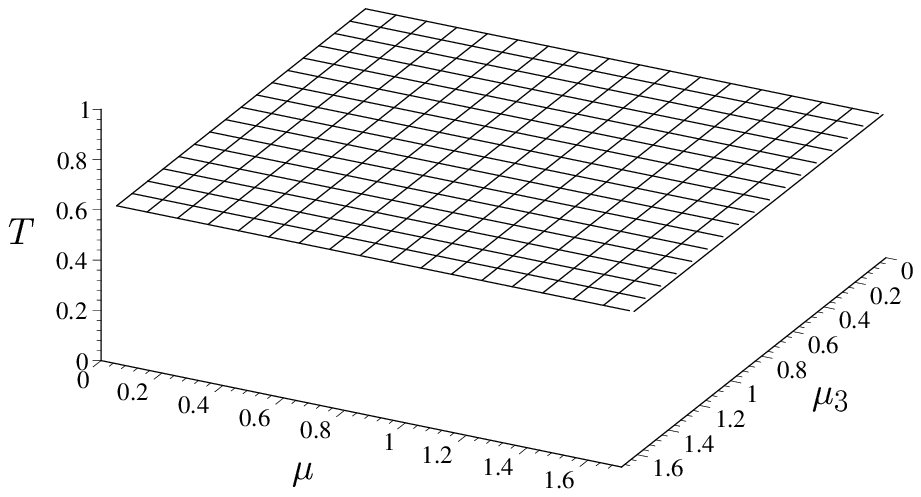,width=6cm,angle=90}
\caption{Phase diagram of the U(2)-NJL model as a function of $\mu,\mu_3,T$. Above the horizontal, shaded surface, chiral symmetry is restored and
the order parameter vanishes. Below the surface, chiral symmetry is broken. The mean field $\Delta$ has the form given in Eq.~(\ref{E6}) corresponding to
two chiral spirals of radius $m(T)$ and different pitches for isospin up and isospin down fermions.}
\label{fig1}
\end{center}
\end{figure}

Without detailed calculation, we can predict the phase diagram in ($\mu,\mu_3,T)$ space by using the known results for the U(1)-NJL and SU(2)-NJL models
\cite{L4,L16,L19}.
For temperatures above $T_c= e^{\rm C}/\pi$, chiral symmetry is restored ($\Delta=0$) and the fermions are massless for all ($\mu,\mu_3$).
Below $T_c$  the order parameter is expected to be 
\begin{equation}
\Delta = m(T) e^{2i(\mu+\mu_3 \tau_3)x}
\label{E6}
\end{equation}
($m(T)$ is the dynamical fermion mass at $\mu=\mu_3=0$)
with non-vanishing components
\begin{eqnarray}
S_0 & = & + m(T)\cos 2 \mu x \cos 2 \mu_3 x, \quad P_0 = - m(T)\sin 2 \mu x \cos 2 \mu_3 x,
\nonumber \\
S_3 & = & - m(T)\sin 2 \mu x \sin 2 \mu_3 x, \quad P_3 = - m(T)\cos 2 \mu x  \sin 2 \mu_3 x,
\label{E7}
\end{eqnarray}
and the densities and grand canonical potential density
\begin{equation}
\rho(T,\mu,\mu_3) =  \frac{2 N_c \mu}{\pi}, \quad \rho_3(T,\mu,\mu_3) =  \frac{2 N_c \mu_3}{\pi}, \quad \Psi(T,\mu,\mu_3) = \Psi(T,0,0) - \frac{N_c(\mu^2+ \mu_3^2)}{\pi}.
\label{E8}
\end{equation}
In the SU(2)-NJL$_2$ model, we had a spatial modulation of the radius of the chiral spiral. Here the radius is spatially
constant, but the pitch gets modulated. Notice the lack of periodicity of $\Delta$ if $\mu$ and $\mu_3$ are incommensurate.
Thus the phase diagram is extremely simple and closely reminiscent of the U(1)-NJL model, see Fig.~\ref{fig1}, in contrast to the more
complicated phase diagram of the Z$_2$-GN and SU(2)-NJL models \cite{L4,L5,L20}. 
%###########################################################################################################################
%############################################################################################################################

It is worth noting that the type of order parameter (\ref{E6}) has appeared before in the literature as variational ansatz for the SU(2)-NJL model in 3+1 dimensions
\cite{L21}. In our case, the fact that the chiral group is U(2) rather than SU(2) is crucial for this simple ansatz to work quantitatively, see also Ref.~\cite{L4}
for a comparison with the phase diagram of the SU(2)-NJL model in 1+1 dimensions.

%<<<<<<<<<<<<<<<<<<<<<<<<<<<<<<<<<<<<<<<<<<<<<<<<<<<<<<<<<<<<<<<<<<<<<<<<<<<<<<<<<<<<<<<<<<<< <<<<<<<<<<<<<<<<<<<<<<<<<<<<<
%<<<<<<<<<<<<<<<<<<<<<<<<<<<<<<<<<<<<<<<<<<<<<<<<<<<<<<<<<<<<<<<<<<<<<<<<<<<<<<<<<<<<<<<<<<<<<<<<<<<<<<<<<<<<<<<<<<<<<<<<<<
\section{Soliton dynamics: Multicomponent TDHF equation}
\label{sect6}
%<<<<<<<<<<<<<<<<<<<<<<<<<<<<<<<<<<<<<<<<<<<<<<<<<<<<<<<<<<<<<<<<<<<<<<<<<<<<<<<<<<<<<<<<<<<<<<<<<<<<<<<<<<<<<<<<<<<<<<<<<<
%<<<<<<<<<<<<<<<<<<<<<<<<<<<<<<<<<<<<<<<<<<<<<<<<<<<<<<<<<<<<<<<<<<<<<<<<<<<<<<<<<<<<<<<<<<<<<<<<<<<<<<<<<<<<<<<<<<<<<<<<<<

A framework which enables us to solve soliton dynamics in the U(2)-NJL model is available
from the theory of fermionic superfluidity and superconductivity in the form of a general solution of the multicomponent BdG equation \cite{L9}. In a previous work, we
have outlined how to adapt this formalism to the SU(2)-NJL model \cite{L4}. Since it is inconvenient to mix two formalisms
with different conventions, and since details of the symmetries of $\Delta$ and occupation fractions may be somewhat different
in condensed matter and particle physics, we proceed in this section to generalize the one-flavor formalism of Refs.~\cite{L13,L14,L15} to two flavors. We are strongly 
guided by Takahashi's work and will arrive at results equivalent to his results, but in such a way that the equations
resemble the previous ones from the one-flavor case and where covariance is more manifest.
We hope that this will make applications to relativistic quantum field theories easier.

After inspecting Takahashi's formalism, one quickly discovers that the generalization
of the framework of Refs.~\cite{L13,L14,L15} is rather straightforward. Due to the additional flavor index,
the notation becomes more cumbersome, otherwise the whole scheme remains practically unchanged.
This is true notably for the construction of transparent potentials where we follow closely the logic
of Ref.~\cite{L14}.

In \cite{L14}, the starting point for attacking $N$ soliton problems was a $N$-dimensional vector $e$ 
with components
\begin{equation}
e_n = e^{i (\zeta_n^* \bar{z} -z/\zeta_n^*)/2}.
\label{F1}
\end{equation}
The $\zeta_n$ are complex numbers (Im $\zeta_n >0$) characterizing the pole positions of the TDHF continuum
wave functions in the complex $\zeta$ plane, $\zeta$ being the spectral parameter
related to light cone momentum and energy (uniformizing parameter in condensed matter language), 
\begin{equation}
k=\frac{1}{2} \left( \zeta- \frac{1}{\zeta}\right), \quad E= - \frac{1}{2} \left( \zeta + \frac{1}{\zeta} \right).
\label{F2}
\end{equation}
Note that
\begin{equation}
k_{\mu}x^{\mu} = - \frac{1}{2} \left( \zeta \bar{z} - \frac{z}{\zeta} \right),
\label{F3}
\end{equation}
so that $e_n$ is recognized as a plane wave evaluated at a complex spectral parameter corresponding to a bound state pole.
The main generalization when going from one to two flavors consists in introducing two copies of each $e_n$, 
\begin{equation}
e_n \to e_n \vec{p}_n , \quad e_{n,\alpha} = e_n p_{n,\alpha}.
\label{F4}
\end{equation}
Here, $\vec{p}_n$ is a 2-component, constant, complex vector with components $p_{n,\alpha}$. 
Its precise meaning will be clarified later on when we investigate few-soliton problems in more detail.
The generalization to more than 2 flavors is straightforward, but not needed here.
Without loss of generality, we can assume that the vectors $\vec{p}_n$ are normalized ($\vec{p}_n^{\, \, \dagger} \vec{p}_n=1$). 
We shall use Greek indices and the 
summation convention for flavor ($\alpha=1...2$) and suppress the indices $n=1...N$ referring to the bound state poles whenever possible.
Then we have to modify the equations of Refs.~\cite{L14,L15} as follows:
Continuum TDHF spinors are now 4-component objects represented as 
\begin{equation}
\psi_{\zeta,\alpha} = \frac{1}{\sqrt{1+\zeta^2}} \left( \begin{array}{c} \zeta \chi_{1,\alpha} \\ - \chi_{2,\alpha} \end{array} \right) e^{i(\zeta \bar{z}-z/\zeta)/2}.
\label{F5}
\end{equation}
The following ansatz is inspired by the assumed pole structure of the continuum spinors ($N$ poles, corresponding to $N$ bound states)
\begin{eqnarray}
\chi_{1,\alpha} & = & \left( \delta_{\alpha \beta} + i \sum_{n=1}^N \frac{1}{\zeta-\zeta_n} e_{n,\beta}^* \varphi_{1,n,\alpha} \right) q_{\beta},
\nonumber \\
\chi_{2,\alpha} & = & \left( \delta_{\alpha \beta} - i \sum_{n=1}^N \frac{\zeta}{\zeta-\zeta_n} e_{n,\beta}^* \varphi_{2,n,\alpha} \right) q_{\beta}.
\label{F6}
\end{eqnarray}
The $q_{\beta}$ are the amplitudes of the flavor components of the incoming plane wave
\begin{equation}
\psi_{\zeta,\alpha}|_{\rm in} = \frac{1}{\sqrt{1+\zeta^2}} \left( \begin{array}{c} \zeta \\ -1 \end{array} \right) e^{i(\zeta \bar{z}- z/\zeta)/2} q_{\alpha}.
\label{F7}
\end{equation}
When summing over all continuum states, they should be chosen as $(q_1,q_2)=(1,0)$ and $(0,1)$ to account for incoming waves in the two
isospin channels (at least, if the vacuum at $x\to -\infty$ is chosen as $\Delta_-=1$).
The $\varphi_{i,n,\alpha}$ are closely related to bound state wave functions. They can be evaluated by linear algebra as follows:
Define a hermitean $N \times N$ matrix $B$, 
\begin{equation}
B_{n m}= i \frac{e_{n,\beta} e_{m,\beta}^*}{\zeta_m-\zeta_n^*} = i \frac{e_n e_m^*}{\zeta_m-\zeta_n^*} \vec{p}_m^{\, \, \dagger}\vec{p}_n .
\label{F8}
\end{equation}
As in Refs.~\cite{L14,L15}, we
suppress soliton indices ($n,m$) to ease the notation.
The $\varphi_{1,n,\alpha},\varphi_{2,n,\alpha}$ satisfy the following system of linear, algebraic equations
\begin{eqnarray}
(\omega + B) \varphi_{1,\alpha} & = & e_{\alpha},
\nonumber \\
(\omega + B) \varphi_{2,\alpha} & = & - f_{\alpha},
\label{F9}
\end{eqnarray}
where $f_{n,\alpha}=e_{n,\alpha}/\zeta_n^*$. 
A  constant, Hermitean $N \times N$ matrix $\omega$ encoding further information
about the soliton configuration (geometry,  initial conditions, details about time dependence for breathers) has been introduced.
The dimension of the linear systems does not increase with the number of flavors, but depends only on the total number of bound state poles.
What is new is the factor $\vec{p}_m^{\, \, \dagger}\vec{p}_n$ in $B_{nm}$ and the fact
that one needs to solve the linear equation for each flavor component $\alpha$ in turn, a rather 
mild complication. Notice that the pairs of exponentials ($e_{n,\alpha}, -f_{n,\alpha}$) provide us with $2 N$ non-normalizable
solutions of the free, massive Dirac equation ($m=1$),
\begin{equation}
2i \partial e_{\alpha} = f_{\alpha}, \quad 2i \bar{\partial} f_{\alpha} = - e_{\alpha}.
\label{F10}
\end{equation}
Derivatives of $B$ are now 2-term separable, as opposed to 1-term separable before,
\begin{equation}
\partial B = \frac{1}{2} f_{\beta} f_{\beta}^{\dagger}, \quad \bar{\partial} B = 
\frac{1}{2} e_{\beta} e_{\beta}^{\dagger}.
\label{F11}
\end{equation}
Differentiating (\ref{F9}) with respect to $z, \bar{z}$ yields
\begin{eqnarray}
(\omega+B) 2i \partial \varphi_{1,\alpha} & = & f_{\beta} \left(\delta_{\alpha,\beta} - i f_{\beta}^{\dagger} \varphi_{1,\alpha}\right),
\nonumber \\
(\omega+B) 2i \bar{\partial} \varphi_{2,\alpha} & = & e_{\beta} \left( \delta_{\alpha \beta} - i e_{\beta}^{\dagger} \varphi_{2, \alpha} \right) .
\label{F12}
\end{eqnarray}
Upon applying $(\omega+B)^{-1}$ from the right, we arrive at 
\begin{eqnarray}
2i \partial \varphi_{1,\alpha} & = &  - \varphi_{2,\beta} \left(\delta_{\alpha \beta} - i f_{\beta}^{\dagger} \varphi_{1,\alpha}\right),
\nonumber \\
2i \bar{\partial} \varphi_{2,\alpha} & = &  \varphi_{1,\beta} \left( \delta_{\alpha \beta} - i e_{\beta}^{\dagger} \varphi_{2, \alpha} \right) .
\label{F13}
\end{eqnarray}
This is just the 2-component version of the covariant TDHF equation with the following 2$\times$2 matrix potential
\begin{equation}
\Delta_{\alpha  \beta} = \delta_{\alpha \beta} - i e_{\beta}^{\dagger} \varphi_{2, \alpha}
= \delta_{\alpha \beta} + i \varphi_{1,\beta}^{\dagger} f_{\alpha}
= \delta_{\alpha \beta}+i e_{\beta}^{\dagger} \frac{1}{\omega+B} f_{\alpha}.
\label{F14}
\end{equation}
Turning to the continuum states, the Dirac equation for $\chi_{1,\alpha}, \chi_{2,\alpha}$ introduced in (\ref{F5}) assumes the form
\begin{eqnarray}
(2i \bar{\partial} - \zeta) \chi_{2,\alpha} + \zeta \Delta_{\alpha \beta} \chi_{1,\beta} & = &  0,
\nonumber \\
(2i  \zeta \partial +1)\chi_{1,\alpha} - \Delta^{\dagger}_{\alpha \beta} \chi_{2,\beta} & = &  0.
\label{F15}
\end{eqnarray}
We write down the three terms of each of these equations in detail. From the first line of (\ref{F15}), 
\begin{eqnarray}
2i \bar{\partial} \chi_{2,\alpha} & = & - i \sum_n \frac{\zeta}{\zeta-\zeta_n} e_{n,\gamma}^* q_{\gamma} \left( \zeta_n \varphi_{2,n,\alpha} 
+ \Delta_{\alpha \beta} \varphi_{1,n,\beta}\right),
\nonumber \\
- \zeta \chi_{2,\alpha} & = & - \zeta q_{\alpha} + i \sum_n \frac{\zeta^2}{\zeta-\zeta_n} e_{n,\gamma}^* q_{\gamma} \varphi_{2,n,\alpha} ,
\nonumber \\
\zeta \Delta_{\alpha \beta} \chi_{1,\beta} & = & \zeta \Delta_{\alpha \beta} q_{\beta} + \zeta \Delta_{\alpha \beta} i \sum_n
\frac{1}{\zeta-\zeta_n} e_{n,\gamma}^* q_{\gamma} \varphi_{1,n,\beta} .
\label{F16}
\end{eqnarray}
As in the one-flavor case, terms containing $\varphi_1$ in the first and 3rd line cancel. Terms containing $\varphi_2$ in the first and 2nd lines
add up to $\zeta(q_{\alpha}-\Delta_{\alpha \beta}q_{\beta})$ and cancel the remaining terms in the 2nd and 3rd line.
Likewise, from the 2nd line of (\ref{F15}),
\begin{eqnarray}
2 i \zeta  \partial \chi_{1,\alpha} & = & -i \sum_n \frac{\zeta}{\zeta-\zeta_n} \left( f_{n,\gamma}^* q_{\gamma} \varphi_{1,n,\alpha} 
+ e_{n,\gamma}^* q_{\gamma} \Delta^{\dagger}_{\alpha \beta} \varphi_{2,n,\beta}\right),
\nonumber \\
\chi_{1,\alpha} & = & q_{\alpha} + i \sum_n \frac{1}{\zeta-\zeta_n} e_{n,\gamma}^* q_{\gamma} \varphi_{1,n,\alpha} ,
\nonumber \\
- \Delta^{\dagger}_{\alpha \beta} \chi_{2,\beta} & = & - \Delta^{\dagger}_{\alpha \beta} q_{\beta} + i \Delta^{\dagger}_{\alpha \beta}
 \sum_n \frac{\zeta}{\zeta-\zeta_n} e_{n,\gamma}^* q_{\gamma} \varphi_{2,n,\beta} .
\label{F17}
\end{eqnarray}
Cancellations work exactly as for one flavor: Terms containing $\varphi_2$ in the 1st and 3rd lines cancel. Terms 
containing $\varphi_1$ in the 1st and 2nd lines add up to $\Delta^{\dagger}_{\alpha \beta}q_{\beta}  - q_{\alpha}$ and cancel the
remaining terms in 2nd and 3rd lines.

Next we consider the question of normalization and orthogonality of the bound state spinors. Using
\begin{eqnarray}
\varphi_{1,n,\alpha} & = & (\omega+B)^{-1}_{nm} e_{m,\alpha},
\nonumber \\
\varphi_{2,n,\alpha} & = & - (\omega+B)^{-1}_{nm} f_{m,\alpha},
\label{F18}
\end{eqnarray}
we show that
\begin{eqnarray}
\varphi_{n,\alpha}^{\dagger} \varphi_{m,\alpha} & = &  \varphi_{1,n,\alpha}^* \varphi_{1,m,\alpha} + \varphi_{2,n,\alpha}^* \varphi_{2,m,\alpha}
\nonumber \\
& = & \left( \frac{1}{\omega+B} (e_{\alpha} e_{\alpha}^{\dagger} + f_{\alpha} f_{\alpha}^{\dagger} )\frac{1}{\omega+B} \right)_{mn}
\nonumber \\
& = & -2 \partial_x \left( \frac{1}{\omega+B} \right)_{mn}.
\label{F19}
\end{eqnarray}
This yields the same overlap matrix as in the one flavor case,
\begin{equation}
R_{nm} = \int_{-\infty}^{\infty} dx \varphi_{n,\alpha}^{\dagger} \varphi_{m,\alpha} = 2 \left(\omega^{-1}\right)_{mn}.
\label{F20}
\end{equation}
Orthonormal bound states can be constructed as before by linear combinations of the $\varphi_n$,
\begin{equation}
\hat{\varphi}_n = \sum_m\ C_{nm} \varphi_m, \quad \int dx \hat{\varphi}_{n \alpha}\hat{\varphi}_{m,\alpha} = \delta_{n,m}.
\label{F21}
\end{equation}
The resulting condition coincides with the one in the one-flavor case,
\begin{equation}
2 C \omega^{-1} C^{\dagger} = 1.
\label{F22}
\end{equation}
We now turn to the self-consistency condition. The mean field $\Delta_{\alpha \beta}$ receives
contributions from the sea and the bound states,
\begin{equation}
\Delta_{\alpha \beta} = - 2 Ng^2 \left( \langle \psi_{2,\alpha} \psi_{1,\beta}^* \rangle_{\rm sea} +  \langle \psi_{2,\alpha} \psi_{1,\beta}^* \rangle_{\rm b}\right) \quad (N=2N_c),
\label{F23}
\end{equation}
with
\begin{eqnarray}
\langle  \psi_{2, \alpha} \psi_{1,\beta}^* \rangle_{\rm sea} & = & -\frac{1}{2} \int_{1/\Lambda}^{\Lambda} \frac{d\zeta}{2\pi} \frac{1}{\zeta}  \chi_{2,\alpha} \chi_{1,\beta}^*,
\nonumber \\
\langle  \psi_{2, \alpha} \psi_{1,\beta}^* \rangle_{\rm b} & = & \sum_{n} \nu_n \hat{\varphi}_{2,n,\alpha} \hat{\varphi}_{1,n,\beta}^* .
\label{F24}
\end{eqnarray}
To evaluate the continuum part, we insert the $\chi$'s and integrate over $d\zeta$ with a cutoff. The pole at $\zeta=0$ yields the divergent contribution
\begin{equation}
\left. \langle \psi_{2,\alpha}   \psi_{1,\beta}^*  \rangle_{\rm sea}\right|_{\rm div} =  - \frac{\Delta_{\alpha \beta}}{2\pi} \ln \Lambda.
\label{F25}
\end{equation}
Owing to the  vacuum gap equation, this part gives self-consistency by itself, as usual in many variants of GN models.
The convergent part  of the sea contribution can be written down most easily if one introduces a diagonal matrix $M$,
\begin{equation}
M_{nm} = -i \delta_{nm} \ln(-\zeta_n^*).
\label{F26}
\end{equation} 
We find
\begin{equation}
\left. \langle \psi_{2,\alpha} \psi_{1, \beta}^* \rangle_{\rm sea}\right|_{\rm conv} = - \frac{1}{4\pi} \varphi_{1,\beta}^{\dagger}  \left( \omega M^{\dagger} +
M \omega \right) \varphi_{2,\alpha}.  
\label{F27}
\end{equation}
The bound state contribution, expressed through the diagonal matrix $N$ with matrix elements proportional to bound state occupation fractions,
\begin{equation}
N_{nm} = 4 \pi \delta_{nm} \nu_n,
\label{F28}
\end{equation}
is given by
\begin{equation}
\langle  \psi_{2,\alpha} \psi_{1,\beta}^* \rangle_{\rm b} =   \frac{1}{4\pi} \varphi_{1,\beta}^{\dagger} \left( C^{\dagger} N C  \right)  \varphi_{2,\alpha}.
\label{F29}
\end{equation}
Expressions (\ref{F27}) and (\ref{F29}) cancel provided that  
\begin{equation}
\omega M^{\dagger} + M \omega = C^{\dagger} N C.
\label{F30}
\end{equation}
Remarkably, the self-consistency condition is not affected at all when going from one to two flavors.

As outlined in Sect.~\ref{sect2},
due to current conservation, the solitons described by this formalism should have identically vanishing baryon density
($\rho = \langle \psi^{\dagger} \psi \rangle$) and isospin densities ($\rho_a = \langle  \psi^{\dagger} \tau_a \psi \rangle $).
This should hold for left- and right-handed fermions separately, or, equivalently, for charge and current densities.
It is a good test of the formalism to prove this in all generality.
Define
\begin{equation}
\rho_L^a  =  \langle \psi_L^{\dagger} \tau^a \psi_L \rangle, \quad \rho_R^a = \langle \psi_R^{\dagger} \tau^a \psi_R \rangle.
\label{F31}
\end{equation}
Introducing $\tau^0=1$ in addition to the ordinary SU(2) generators $\vec{\tau}$, we can treat the baryon density and isovector densities
on the same footing. The densities comprise a part from the Dirac sea and a part from the bound states. The sea part has to be subtracted
by the vacuum contribution for the case $a=0$ only (baryon density). The sea contribution is evaluated as follows. We take into account the
measure $(\zeta^2+1)/2\zeta^2$ and write the total density as the integral
\begin{eqnarray}
\rho_{{\rm sea},L}^a & = &  \int_0^{\infty} \frac{d\zeta}{2\pi} (\tau^a)_{\alpha \beta} X_{L,\alpha \beta},
\nonumber \\
\rho_{{\rm sea},R}^a & = & \int_0^{\infty}  \frac{d\zeta}{2\pi} (\tau^a)_{\alpha \beta} X_{R,\alpha \beta}.
\label{F32}
\end{eqnarray}
Here, $X_{L,R}$ are the contribution from the continuum state with spectral parameter $\zeta$ to the densities, including the measure. Inserting
the continuum spinors
and summing over the two isospin channels, we find
\begin{eqnarray}
X_{L,\alpha \beta} & = & \frac{1}{2} \left( \varphi_{1,\alpha}^{\dagger} g_{\gamma}  g_{\gamma}^{\dagger} \varphi_{1,\beta} -i \varphi_{1,\alpha}^{\dagger} g_{\beta}
+ i g_{\alpha}^{\dagger} \varphi_{1,\beta} \right) ,
\nonumber \\
X_{R,\alpha \beta} & = & \frac{1}{2} \left( \varphi_{2,\alpha}^{\dagger} g_{\gamma}  g_{\gamma}^{\dagger} \varphi_{2,\beta} + \frac{i}{\zeta} \varphi_{2,\alpha}^{\dagger} g_{\beta}
- \frac{i}{\zeta} g_{\alpha}^{\dagger} \varphi_{1,\beta} \right). 
\label{F33}
\end{eqnarray}
We have introduced the vector $g_{\gamma}$ with $N$ components 
\begin{equation}
g_{n,\gamma} = \frac{e_{n,\gamma}}{\zeta- \zeta_n},
\label{F34}
\end{equation}
generalizing the corresponding vector $g$ in the one-flavor case. Consider the quadratic terms in $\varphi_1,\varphi_2$ in Eq.~(\ref{F33}) first. For $X_L$, use the following identity:
\begin{eqnarray}
\varphi_{1,\alpha}^{\dagger} g_{\gamma} g_{\gamma}^{\dagger} \varphi_{1,\beta}  & = &  \varphi_{1,n,\alpha}^* g_{n,\gamma} g_{m,\gamma}^* \varphi_{1,m,\beta}
\nonumber \\
& = & i \varphi_{1,n,\alpha}^* \left( \frac{1}{\zeta-\zeta_n^*}B_{nm}- B_{nm} \frac{1}{\zeta-\zeta_m} \right)  \varphi_{1,m,\beta} .
\label{F35}
\end{eqnarray}
Eq.~(\ref{F9}) in the form 
\begin{equation}
B \varphi_{1,\beta} = e_{\beta} - \omega \varphi_{1,\beta}, \quad \varphi_{1,\beta}^{\dagger} B = e_{\beta}^{\dagger} - \varphi_{1,\beta}^{\dagger} \omega 
\label{F36}
\end{equation}
then serves to eliminate $B_{nm}$ from (\ref{F35}), thereby generating two terms which cancel exactly the terms linear in $\varphi_1$ in (\ref{F33}). The remainder
yields
\begin{equation}
X_{L,\alpha \beta} = \frac{-i }{2}\varphi_{1,n,\alpha}^* \left( \frac{1}{\zeta-\zeta_n^*} \omega_{nm} - \omega_{nm} \frac{1}{\zeta-\zeta_m} \right) \varphi_{1,m,\beta},   
\label{F37}
\end{equation} 
or, with the help of the diagonal matrix
\begin{equation}
{\cal Z} = {\rm diag} (\zeta_1, ..., \zeta_n), 
\label{F38}
\end{equation}
more compactly 
\begin{equation}
X_{L,\alpha \beta}  =  -\frac{i}{2} \varphi_{1,\alpha}^{\dagger} \left( \frac{1}{\zeta-{\cal Z}^{\dagger}} \omega-\omega \frac{1}{\zeta-{\cal Z}} \right) \varphi_{1,\beta}
\label{F39}
\end{equation}
Along the same lines, one finds for the right-handed density
\begin{equation}
X_{R,\alpha \beta}  =  -\frac{i}{2} \varphi_{2,\alpha}^{\dagger} \left( \frac{1}{\zeta-{\cal Z}^{\dagger}} \omega-\omega \frac{1}{\zeta-{\cal Z}} \right) \varphi_{2,\beta}
\label{F40}
\end{equation}
The bound state contribution in a notation similar to (\ref{F31}) is
\begin{eqnarray}
\rho_{b,L}^a  & = &  \sum_n \nu_n \tau_{\alpha \beta}^a \hat{\varphi}_{1,n,\alpha}^* \hat{\varphi}_{1,n,\beta} 
\nonumber \\
\rho_{b,R}^a  & = &   \sum_n \nu_n \tau_{\alpha \beta}^a \hat{\varphi}_{2,n,\alpha}^* \hat{\varphi}_{2,n,\beta}
\label{F41}
\end{eqnarray}
Since
\begin{equation}
 \sum_n \nu_n \tau_{\alpha \beta}^a \hat{\varphi}_{i,n,\alpha}^* \hat{\varphi}_{i,n,\beta} = \frac{1}{4\pi} \tau^a_{\alpha \beta}\varphi_{i,\alpha}^{\dagger} C^{\dagger} N C \varphi_{i,\beta}
\label{F42}
\end{equation}
with the matrix $N$ defined in (\ref{F28}), the condition which guarantees vanishing of all 4 left- and right-handed densities is
\begin{equation}
\frac{1}{4\pi} C^{\dagger} N C -\frac{i}{2} \int_0^{\infty} \frac{d\zeta}{2\pi} \left( \frac{1}{\zeta-{\cal Z}^{\dagger}}\omega - \omega \frac{1}{\zeta- {\cal Z}} \right)  = 0
\label{F43}
\end{equation}
Upon performing the integration over $d\zeta$ and noting that
\begin{equation}
M = - i \ln \left(-{\cal Z}^{\dagger} \right),
\label{F44}
\end{equation} 
we recover the self-consistency condition (\ref{F30}). This shows that the physical solutions indeed have
identically vanishing densities. The result is non-trivial in the sense that it comes about as a cancellation between non-vanishing contributions from
the Dirac sea and the valence bound states and only holds for self-consistent solutions. Hence it may be viewed as a complete shielding effect

%<<<<<<<<<<<<<<<<<<<<<<<<<<<<<<<<<<<<<<<<<<<<<<<<<<<<<<<<<<<<<<<<<<<<<<<<<<<<<<<<<<<<<<<<<<<< <<<<<<<<<<<<<<<<<<<<<<<<<<<<<
%<<<<<<<<<<<<<<<<<<<<<<<<<<<<<<<<<<<<<<<<<<<<<<<<<<<<<<<<<<<<<<<<<<<<<<<<<<<<<<<<<<<<<<<<<<<<<<<<<<<<<<<<<<<<<<<<<<<<<<<<<<
\section{One pole -- the fundamental twisted kink}
\label{sect7}
%<<<<<<<<<<<<<<<<<<<<<<<<<<<<<<<<<<<<<<<<<<<<<<<<<<<<<<<<<<<<<<<<<<<<<<<<<<<<<<<<<<<<<<<<<<<<<<<<<<<<<<<<<<<<<<<<<<<<<<<<<<
%<<<<<<<<<<<<<<<<<<<<<<<<<<<<<<<<<<<<<<<<<<<<<<<<<<<<<<<<<<<<<<<<<<<<<<<<<<<<<<<<<<<<<<<<<<<<<<<<<<<<<<<<<<<<<<<<<<<<<<<<<<

Choose $N=1$ (one soliton) in the general formalism and work it out. After setting
\begin{equation}
\omega_{11} = 1, \quad V_1 = i \frac{|e_1|^2}{(\zeta_1-\zeta_1^*)}
\label{G1}
\end{equation} 
we find the mean field
\begin{equation}
\Delta = 1 + \left( \frac{\zeta_1}{\zeta_1^*}-1 \right) \frac{V_1}{1+V_1} \vec{p}_1 \vec{p}_1^{\,\dagger}.
\label{G2}
\end{equation}
This is the most elementary type of kink interpolating between two vacua. In the rest frame,
\begin{equation}
\zeta_1 = - e^{-i\phi_1}, \quad V_1 = e^{2x \sin \phi_1}.
\label{G3}
\end{equation}
For $x\to -\infty$, $\Delta=\Delta_- = 1$. For $x\to \infty$, $\Delta = \Delta_+$ with
\begin{equation}
\Delta_+ = 1 + \left( e^{-2i\phi_1}-1 \right) \vec{p}_1 \vec{p}_1^{\, \dagger}.
\label{G4}
\end{equation}
By a unitary transformation we can map $\vec{p}_1$ onto the vector $(1,0)^T$, in which case we recover
the standard U(1)-NJL kink in the isospin-up channel and the vacuum $\Delta_+$ in the diagonal form
\begin{equation}
\Delta_+ = \left( \begin{array}{cc} e^{-2i\phi_1} & 0 \\ 0 & 1 \end{array} \right).
\label{G5}
\end{equation}
Hence we have a twisted kink in the isospin up channel and the vacuum for isospin down.
The general expression (\ref{G4}) is nothing but the spectral representation of the unitary matrix $\Delta_+$.
Denoting the (normalized) eigenvectors corresponding to the eigenvalues $e^{-2i\phi_1}$ and 1 by 
$\vec{p}_1$ and $\vec{p}_2$, respectively, we have
\begin{equation}
\Delta_+ =  e^{-2i\phi_1} \vec{p}_1 \vec{p}_1^{\, \dagger} + \vec{p}_2 \vec{p}_2^{\, \dagger}.
\label{G6}
\end{equation}
Eliminating the second term with the help of the completeness relation
\begin{equation}
\vec{p}_1 \vec{p}_1^{\, \dagger} +  \vec{p}_2 \vec{p}_2^{\, \dagger}= 1
\label{G7}
\end{equation}
then reproduces (\ref{G4}). This shows that the vector $\vec{p}_1$ can be interpreted as follows: $\vec{p}_1$ is the
eigenvector of the unitary matrix $\Delta_+$, the vacuum at $x\to \infty$, belonging to the ``twisted" eigenvalue
$e^{-2i\phi_1}$. With one bound state, only one eigenvalue can be twisted. This is the fundamental twisted kink.

Since we can reduce the simplest kink problem to the kink of the U(1)-NJL model,   
the spinors, self-consistency and vanishing baryon and isospin density follow
from one-flavor results and need not be repeated here.

This is exactly the kink of Takahashi \cite{L9}. It is special in the following sense: 
A generic twisted kink in the U(2)-NJL model should connect the vacuum $\Delta_-=1$ with the vacuum $\Delta_+$, a 
unitary matrix with eigenvalues $e^{-2i\phi_1}, e^{-2i\phi_2}$. In the present framework this requires
a bound state of two elementary kinks, see next section. This was already used in a previous work
on the SU(2)-NJL model \cite{L4}, where the eigenvalues have to be complex conjugates $e^{\mp 2i\phi_1}$.
It is then necessary to bind a kink with its charge conjugate. No such restriction exists in the U(2)-NJL model.
The elementary kink exists as a physical state in the U(2)-NJL model, whereas it is confined in the
SU(2)-NJL model. This is completely analoguous to what happens with Shei's twisted kink \cite{L11} in the U(1)-NJL and  Z$_2$-GN models, 
respectively \cite{L15}.

%<<<<<<<<<<<<<<<<<<<<<<<<<<<<<<<<<<<<<<<<<<<<<<<<<<<<<<<<<<<<<<<<<<<<<<<<<<<<<<<<<<<<<<<<<<<< <<<<<<<<<<<<<<<<<<<<<<<<<<<<<
%<<<<<<<<<<<<<<<<<<<<<<<<<<<<<<<<<<<<<<<<<<<<<<<<<<<<<<<<<<<<<<<<<<<<<<<<<<<<<<<<<<<<<<<<<<<<<<<<<<<<<<<<<<<<<<<<<<<<<<<<<<
\section{Two poles -- scattering of fundamental kinks, generic kink and breather}
\label{sect8}
%<<<<<<<<<<<<<<<<<<<<<<<<<<<<<<<<<<<<<<<<<<<<<<<<<<<<<<<<<<<<<<<<<<<<<<<<<<<<<<<<<<<<<<<<<<<<<<<<<<<<<<<<<<<<<<<<<<<<<<<<<<
%<<<<<<<<<<<<<<<<<<<<<<<<<<<<<<<<<<<<<<<<<<<<<<<<<<<<<<<<<<<<<<<<<<<<<<<<<<<<<<<<<<<<<<<<<<<<<<<<<<<<<<<<<<<<<<<<<<<<<<<<<<

We start with the scattering of two elementary kinks of the type discussed in the preceding section. This requires a diagonal matrix $\omega$. We set
\begin{equation}
\omega_{11} = \omega_{22} = 1, \quad V_1 = i \frac{|e_1|^2}{(\zeta_1-\zeta_1^*)}, \quad \xi V_2 = i \frac{|e_2|^2}{(\zeta_2-\zeta_2^*)}
\label{H1}
\end{equation}
where $\xi$ will be interpreted below.
Then the result of using the general formalism for two poles ($N=2$) follows the same pattern as kink-kink scattering in the U(1)-NJL model or SU(2)-NJL model.
The mean field can be cast into the form
\begin{equation}
\Delta = \frac{1+ U_1V_1 + U_2 \xi V_2 + U_{12} V_1 V_2}{1+V_1+\xi V_2 + V_1V_2}.
\label{H2}
\end{equation}
The interpretation of the U(2) matrices $U_1,U_2,U_{12}$ and of the factor $\xi$ follows upon considering the asymptotics of the scattering
process, 
\begin{eqnarray}
\Delta_{\rm 1,in} & = & \lim_{V_2 \to 0} \Delta = \frac{1+U_1 V_1}{1+V_1},
\nonumber \\
\Delta_{\rm 2,in} & = & \lim_{V_1 \to \infty} \Delta = \frac{1+ U_{12}U_1^{\dagger} V_2}{1+V_2}U_1,
\nonumber \\
\Delta_{\rm 1,out} & = & \lim_{V_2 \to \infty} \Delta = \frac{1 +  U_{12}U_2^{\dagger} \xi^{-1} V_1}{1+\xi^{-1} V_1}U_2,
\nonumber \\
\Delta_{\rm 2,out} & = & \lim_{V_1 \to 0} \Delta = \frac{1+U_2 \xi V_2}{1+ \xi V_2}.
\label{H3}
\end{eqnarray}
Thus $\xi$ accounts for the time delay of the solitons during the collision, and we have made explicit the intrinsic form of the soliton by 
pulling out the factors $U_1$ (for $\Delta_{\rm 2,in}$) and $U_2$ (for $\Delta_{\rm 1,out}$) to the right. One recognizes  $U_1, U_2, U_{12}U_1^{\dagger}$
and $U_{12}U_2^{\dagger}$ as the intrinsic twist factors of the participating solitons. The formalism yields the following results
\begin{eqnarray}
U_1 & = & 1 + \frac{\zeta_1-\zeta_1^*}{\zeta_1^*} \vec{p}_1 \vec{p}_1^{\,\dagger},
\nonumber \\
U_{12} U_1^{\dagger} & = & 1 + \frac{\zeta_2-\zeta_2^*}{\zeta_2^*} \vec{q}_2 \vec{q}_2^{\,\dagger},
\nonumber \\
U_{12} U_2^{\dagger}  & = &   1 + \frac{\zeta_1-\zeta_1^*}{\zeta_1^*} \vec{q}_1 \vec{q}_1^{\,\dagger},
\nonumber \\
U_2 & = & 1 + \frac{\zeta_2-\zeta_2^*}{\zeta_2^*} \vec{p}_2 \vec{p}_2^{\,\dagger}.
\label{H4}
\end{eqnarray}
In this formulation, the normalized vectors $\vec{p}_1, \vec{p}_2$ are input and specify the twist of incoming kink 1 and outgoing kink 2.
The normalized vectors $\vec{q}_1, \vec{q}_2$ of the twist of outgoing kink 1 and incoming kink 2 can be expressed 
in terms of $\vec{p}_1, \vec{p}_2$ and the pole positions $\zeta_i$ as follows,
\begin{eqnarray}
\vec{q}_1 & = & {\cal N}_1 \left[  \sigma_{21} \zeta_1^* (\zeta_2-\zeta_2^*) \vec{p}_2 - \zeta_2^* (\zeta_2-\zeta_1^*) \vec{p}_1 \right],
\nonumber \\
\vec{q}_2 & = & {\cal N}_2 \left[  \sigma_{12} \zeta_2^* (\zeta_1-\zeta_1^*) \vec{p}_1 - \zeta_1^* (\zeta_1-\zeta_2^*) \vec{p}_2 \right].
\label{H5}
\end{eqnarray}
We have used the notation
\begin{equation}
\sigma_{12} = \vec{p}_1^{\,\dagger} \vec{p}_2, \quad \sigma_{21} = \vec{p}_2^{\,\dagger} \vec{p}_1 = \sigma_{12}^*.
\label{H6}
\end{equation}
The normalization factors ensuring that $\vec{q}_1^{\,\dagger}\vec{q}_1 = \vec{q}_2^{\,\dagger} \vec{q}_2 = 1$ are given by
\begin{eqnarray}
{\cal N}_1 & = & \frac{\sqrt{\xi}}{|\zeta_2 (\zeta_2-\zeta_1^*)|},
\nonumber \\
{\cal N}_2 & = & \frac{\sqrt{\xi}}{|\zeta_1 (\zeta_2-\zeta_1^*)|}.
\label{H7}
\end{eqnarray}
Here, $\xi$ is the time delay factor also appearing in Eqs.~(\ref{H2},\ref{H3}) for which we find 
\begin{equation}
\xi^{-1} =  1- \frac{(\zeta_1-\zeta_1^*)(\zeta_2-\zeta_2^*)}{(\zeta_1-\zeta_2^*)(\zeta_2-\zeta_1^*)} \sigma_{12} \sigma_{21}.
\label{H8}
\end{equation}
Let us consider some interesting 
limiting cases. Obviously, the modulus of $\sigma_{12}$ controls the strength of the kink-kink interaction between 
the two kinks. If $\vec{p}_1, \vec{p}_2$ are parallel, $|\sigma_{12}|=1$ and everything can be rotated into the isospin up channel where we recover
scattering of two standard kinks of the U(1)-NJL model, with the known time delay factor
\begin{equation}
\xi = \left| \frac{\zeta_1-\zeta_2^*}{\zeta_1-\zeta_2} \right|^2.
\label{H9}
\end{equation}
If $\vec{p}_1, \vec{p}_2$ are orthogonal, $\sigma_{12}=0$ and the two kinks live in the two isospin channels without any interaction ($\xi=1$).
By varying the angle spanned by $\vec{p}_1$ and $\vec{p}_2$, we can thus vary the strength of the interaction between these two extreme cases.

A special case of the kink-kink scattering problem is the bound state. This may be viewed as the generic U(2)-twisted kink, since it 
enables us to connect an arbitrary vacuum $\Delta_+$ at $x\to \infty$ to $\Delta_-=1$ at $x\to -\infty$. In the rest frame of such a
composite kink, we have to choose
\begin{equation}
\zeta_1 = - e^{-i\phi_1}, \quad \zeta_2 =- e^{-i\phi_2}
\label{H10}
\end{equation}
Some simplifications occur,
\begin{eqnarray}
\vec{q}_1 & = & {\cal N}_1\left[ \left( e^{i(\phi_1+\phi_2)}-1\right) \vec{p}_1 - 2 i  \sigma_{21} e^{i\phi_1} \sin \phi_2 \vec{p}_2\right]
\nonumber \\
\vec{q}_2 & = & - {\cal N}_2 \left[ 2i \sigma_{12} e^{i\phi_2} \sin \phi_1 \vec{p}_1 - \left( e^{i(\phi_1+\phi_2)}-1 \right) \vec{p}_2 \right]
\nonumber \\
{\cal N}_1^{-2} &=& {\cal N}_2^{-2} = 2 \left( 1- \cos (\phi_1+\phi_2) - 2 \sigma_{12} \sigma_{21} \sin \phi_1 \sin \phi_2 \right)
\nonumber \\
\xi^{-1} & = & 1 - \sigma_{12} \sigma_{21} \frac{2 \sin \phi_1 \sin \phi_2}{1-\cos(\phi_1+\phi_2)}
\label{H11}
\end{eqnarray}
In the one flavor case, it was not possible to construct a bound state of two kinks with the same twist, $\phi_1=\phi_2$. If one specializes the 
formalism to this case, one recovers a single kink. In the two flavor case, this restriction does not exist anymore and we can construct a non-trivial
bound state out of two constituents with the same twist. In this case, Eq.~(\ref{H11}) reduces to
\begin{eqnarray}
\vec{q}_1 & = &  {\cal N}_1 \left( e^{2i \phi_1}-1 \right) \left(\vec{p}_1 -  \sigma_{21}\vec{p}_2  \right)
\nonumber \\
\vec{q}_2 & = &  - {\cal N}_2 \left( e^{2i\phi_1}-1 \right) \left( \sigma_{12} \vec{p}_1 - \vec{p}_2 \right)
\nonumber \\
{\cal N}_1^{-2} & = & {\cal N}_2^{-2} = 4 \sin^2 \phi_1 \left(1-\sigma_{12} \sigma_{21} \right)
\nonumber \\
\xi^{-1} & = & 1- \sigma_{12}\sigma_{21}
\label{H12}
\end{eqnarray} 
Finally, we turn to the twisted breather. A breather at rest can be generated by choosing $\eta_1=\eta_2=1$ and a non-diagonal matrix $\omega$.
The twisted breather is even more complicated here than in the single flavor case, so we refrain from discussing all possible parameter choices
and illustrate the reult for a few simple special cases only. We choose the same matrix $\omega$ than for one flavor \cite{L15},
\begin{equation}
\omega = \frac{1}{\cos \chi}\left( \begin{array}{cc} 1 & \sin \chi \\ \sin \chi & 1 \end{array} \right) ,
\label{H13}
\end{equation}
and the vectors $\vec{p}_1, \vec{p}_2$ as
\begin{equation}
\vec{p}_1 = \left( \begin{array}{c} 1 \\ 0 \end{array} \right), \quad \vec{p}_2 = \left( \begin{array}{c} \cos \theta_2 \\ \sin \theta_2 \end{array} \right).
\label{H14}
\end{equation}
Thus $\theta_2$ is the angle between $\vec{p}_1$ and $\vec{p}_2$,
\begin{equation}
\sigma_{12} = \sigma_{21} = \cos \theta_2.
\label{H15}
\end{equation}
If we choose $\vec{p}_1$ and $\vec{p}_2$ to be parallel ($\theta_2=0$), we find that 
\begin{equation}
\Delta = \left( \begin{array}{cc} \Delta_{11} & 0 \\ 0 & 1 \end{array} \right)
\label{H16}
\end{equation}
where $\Delta_{11}$ is the twisted breather in the one-flavor case \cite{L15},
\begin{eqnarray}
\Delta_{11} & = & \frac{{\cal N}_{11}}{\cal D}
\nonumber \\
{\cal N}_{11} & = & 1 + \frac{1}{\cos \chi} \left( \frac{\zeta_1}{\zeta_1^*} V_1 + \frac{\zeta_2}{\zeta_2^*} V_2 \right)+\tan \chi \left( \frac{\zeta_1}{\zeta_2^*}W + \frac{\zeta_2}{\zeta_1^*}W^*\right)
+ \frac{(\zeta_1-\zeta_2)(\zeta_1^*-\zeta_2^*)\zeta_1 \zeta_2}{(\zeta_1^*-\zeta_2)(\zeta_1-\zeta_2^*)\zeta_1^*\zeta_2^*} V_1 V_2
\nonumber \\
{\cal D} & = & 1+ \frac{1}{\cos \chi} \left( V_1+V_2 \right) + \tan \chi \left( W+ W^* \right) +  \frac{(\zeta_1-\zeta_2)(\zeta_1^*-\zeta_2^*)}{(\zeta_1^*-\zeta_2)(\zeta_1-\zeta_2^*)} V_1 V_2
\nonumber \\
V_1 & = & \frac{i |e_1|^2}{\zeta_1-\zeta_1^*}, \quad V_2 = \frac{i |e_2|^2}{\zeta_2-\zeta_2^*}, \quad W = - \frac{i e_1^* e_2}{\zeta_1-\zeta_2^*}
\nonumber \\
\zeta_1 & = & - e^{-i \phi_1}, \quad \zeta_2 = - e^{-i\phi_2}
\label{H17}
\end{eqnarray}
Choosing $\vec{p}_1$ and $\vec{p}_2$ to be orthogonal ($\theta_2=\pi/2$) one finds
\begin{equation}
\Delta = \frac{1}{\cal D} \left( \begin{array}{cc} {\cal N}_{11} & {\cal N}_{12} \\ {\cal N}_{21} & {\cal N}_{22} \end{array} \right)
\label{H18}
\end{equation}
with
\begin{eqnarray}
{\cal N}_{11} & = & 1 + \frac{\zeta_1}{\zeta_1^* \cos \chi} V_1 + \frac{1}{\cos \chi} V_2 + \frac{\zeta_1}{\zeta_1^*} V_1 V_2 
\nonumber \\
{\cal N}_{12} & = & \frac{(\zeta_2-\zeta_1^*)\tan \chi}{\zeta_1^*} W^*
\nonumber \\
{\cal N}_{21} & = & \frac{(\zeta_1-\zeta_2^*)\tan \chi}{\zeta_2^*} W
\nonumber \\
{\cal N}_{22} & = & 1 + \frac{1}{\cos \chi} V_1 + \frac{\zeta_2}{\zeta_2^* \cos \chi} V_2 + \frac{\zeta_2}{\zeta_2^*} V_1 V_2 
\nonumber \\
{\cal D} & = & 1 + \frac{1}{\cos \chi}(V_1+V_2) + V_1 V_2
\label{H19}
\end{eqnarray}
Here, the diagonal components $\Delta_{11}, \Delta_{22}$ are static, whereas the off-diagonal components $\Delta_{12}, \Delta_{21}$ oscillate
with the same frequency as the one-flavor breather,
\begin{equation}
\Omega = \cos \phi_1 - \cos \phi_2.
\label{H20}
\end{equation}
For any other choice of the angle $\theta_2$, all components of $\Delta$ start to oscillate with the same frequency but different phases. The corresponding more complicated
expressions for 
$\Delta$ can easily be generated using the general framework, but will not be given here.

%<<<<<<<<<<<<<<<<<<<<<<<<<<<<<<<<<<<<<<<<<<<<<<<<<<<<<<<<<<<<<<<<<<<<<<<<<<<<<<<<<<<<<<<<<<<< <<<<<<<<<<<<<<<<<<<<<<<<<<<<<
%<<<<<<<<<<<<<<<<<<<<<<<<<<<<<<<<<<<<<<<<<<<<<<<<<<<<<<<<<<<<<<<<<<<<<<<<<<<<<<<<<<<<<<<<<<<<<<<<<<<<<<<<<<<<<<<<<<<<<<<<<<
\section{Summary and conclusions}
\label{sect9}
%<<<<<<<<<<<<<<<<<<<<<<<<<<<<<<<<<<<<<<<<<<<<<<<<<<<<<<<<<<<<<<<<<<<<<<<<<<<<<<<<<<<<<<<<<<<<<<<<<<<<<<<<<<<<<<<<<<<<<<<<<<
%<<<<<<<<<<<<<<<<<<<<<<<<<<<<<<<<<<<<<<<<<<<<<<<<<<<<<<<<<<<<<<<<<<<<<<<<<<<<<<<<<<<<<<<<<<<<<<<<<<<<<<<<<<<<<<<<<<<<<<<<<<

In this paper, we have studied a variant of the integrable Gross-Neveu model family which has not yet received much attention so far in 
1+1 dimensions: The U(2)-NJL model with U(2)$_L \times$U(2)$_R$ chiral symmetry. Phenomenologically, in 3+1 dimensions, the SU(2)$_L \times$SU(2)$_R$
model is more relevant. However, from a theoretical point of view, it is quite instructive to include the U(2) model as well. After developing the mean field approach
and setting up the TDHF equation for the U(2)-NJL model, we have studied the vacuum and identified the vacuum manifold as U(2). Small fluctuations in the 4 flat directions
give rise to 4 massless pseudoscalar mesons, whereas the other 4 directions yield massive scalars right at threshold $({\cal M}=2m)$. The chiral spiral construction is especially
simple here,
as it can be invoked both for baryonic and isospin charge. The phase diagram is extremely simple, generalizing the known U(1)-NJL phase diagram to the
($T,\mu,\mu_3$) space. Perhaps the most interesting topic is that of soliton dynamics. Here it turns out that a recently developed solution of the matrix
BdG system in condensed matter physics fits perfectly the U(2)-NJL model. We have rederived Takahashi's results in the language of our previous work
on the U(1)-NJL model, and confirmed
that the fundamental kink is a physical state in the U(2) model while appearing only as a confined constituent in the SU(2) model. This is another striking
example for the close relationship between relativistic quantum field theory toy models and sophisticated, realistic condensed matter problems. 

In the present work together with Ref.~\cite{L4}, we have generalized the well-known Z$_2$-GN and U(1)-NJL models with Abelian chiral groups to two flavors and
non-Abelian chiral groups SU(2) and U(2). The relationship between these various models is summarized in Table I.
%<<<<<<<<<<<<<<<<<<<<<<<<<<<<<<<<<<<<<<<<<<<<<<<<<<<<<<<<<<<<<<<<<<<<<<<<<<<<<<<<<<<<<<<<<<<<<<<<<<<<<<<<<<<<<<<<<<<<<<<<<<
\begin{table}[h]
\begin{tabular}{|c|c|}
\hline
Z$_2$-GN & U(1)-NJL  \\ 
\hline
SU(2)-NJL & U(2)-NJL 
\\ \hline
\end{tabular}
\caption{Relationship among four-fermion models as discussed in main text.}
\end{table}
%<<<<<<<<<<<<<<<<<<<<<<<<<<<<<<<<<<<<<<<<<<<<<<<<<<<<<<<<<<<<<<<<<<<<<<<<<<<<<<<<<<<<<<<<<<<<<<<<<<<<<<<<<<<<<<<<<<<<<<<<<<
The first row contains the
original variants of the GN model \cite{L1} with one flavor only and discrete or continuous chiral symmetry, respectively. The most conspicuous differences between
these two models are the phase diagram in the ($T,\mu$) plane and the role played by twisted kinks, which are free in the U(1) model but confined into
bound states in the Z$_2$ model. This pattern repeats itself in the 2nd row, the two-flavor generalizations. 
In fact, models in the same column share identical phase diagrams in the ($T,\mu$) plane, as first noticed in Ref.~\cite{L5} for the first column.
A further generalization to the groups SU($N$) and U($N$) is straightforward, thereby extending the spectrum
of integrable quantum field theories substantially. It is plausible (but ought to be checked in future work) that the marked differences in the
phase diagram and in the role of twisted kinks show up for arbitrary numbers of flavors as well.

Our final remark concerns the relationship between the leading order large $N_c$ results and 
the method of non-Abelian bosonization \cite{L22}. It is well known that two-dimensional multi-color and -flavor NJL models
at any finite $N_c, N_f$ can be mapped onto decoupled bosonic field theories of Wess-Zumino-Novikov-Witten type \cite{L22,L23,L24}
and free bosonic fields. Recently there has been a lot of progress in solving such models using conformal field theory
techniques and numerical methods, even for the case when chiral symmetry is explicitly broken by a mass term and
more than one coupling constants, see Ref.~\cite{L25} and references therein. It would be interesting to compute space-time dependent condensates in
these integrable models at finite $N_c,N_f$ and compare them with the results in the limit $N_c \to \infty$,
as this limit is expected to be rather singular.  

%<<<<<<<<<<<<<<<<<<<<<<<<<<<<<<<<<<<<<<<<<<<<<<<<<<<<<<<<<<<<<<<<<<<<<<<<<<<<<<<<<<<<<<<<<<<< <<<<<<<<<<<<<<<<<<<<<<<<<<<<<
%<<<<<<<<<<<<<<<<<<<<<<<<<<<<<<<<<<<<<<<<<<<<<<<<<<<<<<<<<<<<<<<<<<<<<<<<<<<<<<<<<<<<<<<<<<<<<<<<<<<<<<<<<<<<<<<<<<<<<<<<<<
\section*{Acknowledgement}
%<<<<<<<<<<<<<<<<<<<<<<<<<<<<<<<<<<<<<<<<<<<<<<<<<<<<<<<<<<<<<<<<<<<<<<<<<<<<<<<<<<<<<<<<<<<<<<<<<<<<<<<<<<<<<<<<<<<<<<<<<<
%<<<<<<<<<<<<<<<<<<<<<<<<<<<<<<<<<<<<<<<<<<<<<<<<<<<<<<<<<<<<<<<<<<<<<<<<<<<<<<<<<<<<<<<<<<<<<<<<<<<<<<<<<<<<<<<<<<<<<<<<<<
The author thanks Falk Bruckmann for his interest in this work and useful discussions.

%#################################################################################################################################
 
%#######################################################################################################################################


\begin{thebibliography}{99}
\bibitem{L1}
D. J. Gross and A. Neveu, Phys. Rev. D {\bf 10}, 3235 (1974).
\bibitem{L2}
Y. Nambu and G. Jona-Lasinio, Phys. Rev. {\bf 122}, 345 (1961).
\bibitem{L3}
Y. Nambu and G. Jona-Lasinio, Phys. Rev. {\bf 124}, 246 (1961).
\bibitem{L4}
M. Thies, Phys. Rev. D {\bf 93}, 085024 (2016).
\bibitem{L5}
A. Heinz, F. Giacosa, M. Wagner, D. H. Rischke, Phys. Rev. D {\bf 93}, 024512 (2016).
\bibitem{L6}
D. Ebert and M. K. Volkov, Z. Phys. C {\bf 16}, 205 (1983).
\bibitem{L7}
D. Ebert and H. Reinhardt, Nucl. Phys. B {\bf 271}, 188 (1986).
\bibitem{L8}
D. Ebert, M. Nagy, M. K. Volkov, Phys. Atom. Nucl. {\bf 59}, 140 (1996).
\bibitem{L8a}
A. M. Tsvelik, Sov. Phys. JETP {\bf 66}, 754 (1987).
\bibitem{L9}
D. A. Takahashi,  Phys. Rev. B {\bf 93}, 024512 (2016).
\bibitem{L10}
D. A. Takahashi, Prog. Theor. Exp. Phys. {\bf 2016}, 043I01.
\bibitem{L11}
S.-S. Shei, Phys. Rev. D {\bf 14}, 535 (1976).
\bibitem{L12}
R. F. Dashen, B. Hasslacher, A. Neveu, Phys. Rev. D {\bf 12}, 2443 (1975).
\bibitem{L13}
G. V. Dunne and M. Thies, Phys. Rev. D {\bf 89}, 025008 (2014).
\bibitem{L14}
G. V. Dunne and M. Thies, Phys. Rev. Lett. {\bf 111}, 121602 (2013).
\bibitem{L15}
G. V. Dunne and M. Thies, Phys. Rev. A {\bf 88}, 062115 (2013).
\bibitem{L16}
V. Sch\"on and M. Thies, {\em At the frontiers of particle physics: Handbook of QCD, Boris Ioffe Festschrift}, edited by M. Shifman
(World Scientific, Singapore, 2001), Vol. 3, p. 1945.
\bibitem{L17}
R. Pausch, M. Thies, V. L. Dolman, Z. Phys. A {\bf 338}, 441 (1991).
\bibitem{L17a}
W. R. Gutierrez, Nucl. Phys. B {\bf 176}, 185 (1980).
\bibitem{L17b}
M. Cavicchi, P. Di Vecchia, I. Pesando, Mod. Phys. Lett. A {\bf 8}, 2427 (1993); Erratum: Mod. Phys. Lett. A {\bf 8}, 2909 (1993).
\bibitem{L17c}
R. de Mello Koch, J. P. Rodrigues, Phys. Rev. D {\bf 54}, 7794 (1996).
\bibitem{L18}
V. Sch\"on and M. Thies, Phys. Rev. D {\bf 62}, 096002 (2000).
\bibitem{L19}
G. Basar, G. V. Dunne, M. Thies, Phys. Rev. D {\bf 79}, 105012 (2009).
\bibitem{L20}
M. Thies and K. Urlichs, Phys. Rev. D {\bf 67}, 125015 (2003).
\bibitem{L21}
D. Nowakowski, M. Buballa, S. Carignano, J. Wambach, arXiv:1506.04260 [hep-ph].
\bibitem{L22}
E. Witten, Comm. Math. Phys. {\bf 92}, 455 (1984).
\bibitem{L23}
J. Wess and B. Zumino, Phys. Lett. B {\bf 37}, 95 (1971).
\bibitem{L24}
S. P. Novikov, Sov. Math. Dok. {\bf 24}, 222 (1981).
\bibitem{L25}
P. Azaria, R. M. Konik, Ph. Lecheminant, T. P\'almai, G. Tak\'acs, A. M. Tsvelik, arXiv:1601.02979 [hep-th]
\end{thebibliography}
\end{document}